# High Entropy Alloy Nanoparticles Decorated, p-type 2D-Molybdenum Disulphide (MoS$_2$) and Gold Schottky Junction Enhanced Hydrogen Sensing

Kusuma Urs MB[1#], Nirmal Kumar Katiyar[2#], Ritesh Kumar[3], Krishanu Biswas[2*], Abhishek K. Singh[3*], C S Tiwary[4*] and Vinayak Kamble[1*]

[1] School of Physics, Indian Institute of Science Education and Research, Thiruvananthapuram, Kerala -695551 INDIA

[2] Department of Materials Science and Engineering, Indian Institute of Technology Kanpur, Kanpur-208016 INDIA

[3] Materials Research Centre, Indian Institute of Science, Bangalore, Karnataka-560012, INDIA

[4] Metallurgical and Materials Engineering, Indian Institute of Technology Kharagpur, Kharagpur-382355, INDIA

**Keywords:** MoS$_2$ nanosheets, high entropy alloys, Sonochemical, cryomilling, Schottky Junction.

## Abstract

Molybdenum Disulphide (MoS$_2$) is an interesting material which exists in atomically thin multilayers and can be exfoliated into monolayer or a few layers for multiple applications. It has emerged as a promising material for development of such efficient sensors. Here, we have exfoliated and decorated MoS$_2$ flakes with novel, single phase High Entropy Alloy (HEA) nanoparticles using facile and scalable cryomilling technique, followed by sonochemical method. It is found that the decoration of HEA nanoparticles impart surface enhanced Raman scattering effect and reduction in the work function of the material from 4.9 to 4.75 eV as measured by UV photoelectron spectroscopy. This change in the work function results in a schottky barrier between the gold contact and HEA decorated MoS$_2$ flakes as a result of drastic changes in the surface chemical non-stoichiometry. The response to hydrogen gas is studied at temperatures 30 to 100 °C and it shows unusual *p*-type nature due to surface adsorbed oxygen species. The nanoscale junction formed between HEA and MoS$_2$ shows 10 times increase in the response towards hydrogen gas at 80 °C. The experimental observations are further explained with DFT simulation showing more favourable hydrogen adsorption on HEA decorated MoS$_2$ resulting for enhanced response.

#equal contribution
Email: kbvinayak@iisertvm.ac.in, chandra.tiwary@metal.iitkgp.ac.in, kbiswas@iitk.ac.in, abhishek@iisc.ac.in

1. **Introduction**

The high surface area of the atomically thin materials (2D materials) are utilized for different kinds of applications starting from electronics, energy generation, biomedical and sensors[1-6]. Further, the varied and tunable electronic properties of these materials make them a hot cake for various optoelectronic applications. The transition metal dichalcogenides (TMDs) are one of the exciting family of 2D materials, which are reported to be interesting for its varied physical and chemical properties. Particularly in the area of chemical sensors, with the discovery of graphene and TMDs, the onus has been shifted from metal oxide materials to them due to their excellent sensing properties such as sensitivity at relatively low operating temperatures[7-8]. A suitable low operating temperature sensor device, which does not need a power-hungry heater assembly and can make them energy efficient, is required. The high chemical sensitivity of the molybdenum disulphides is utilized for the development of different class of gas sensors[8-10]. However, the major issue of the $MoS_2$ based sensors is their lower sensitivities and large response time i.e., time taken by sensor to reach 90% of the saturation for a given concentration of gas as well as recovery time i.e., time taken to come back to 90% of the initial value upon withdrawing the test gas. This large recovery times arise due to high binding energies of the gas molecules on to the highly reactive $MoS_2$ surface[8, 11]. This could be addressed by making heterostructures of $MoS_2$ with different materials having better electronic properties such as nanoparticles of metals, alloys, oxides, other chalcogenides etc [Ref]. However, the methods of synthesis of such heterostructures are quite laborious, involves multistep. Thus, a simple, facile and scalable method is always desirable.

Recently developed, multicomponent High Entropy Alloys (HEA) have attracted lot of attention due to its unique properties, which are distinctly different than that of its individual components[12-13]. The structural stability and high chemical activity of these HEA has been utilized in structural and energy generation applications. In current work, we have synthesized a hybrid consisting nanoparticles of HEA and $MoS_2$. The HEA nanoparticles are uniformly decorated on the 2D sheet of MoS2. The hybrid is further used for gas sensing properties. In order to gain insights into efficient hydrogen sensing we have performed DFT calculations. It was found that there is a significant structural change at the interface of HEA and $MoS_2$, due to charge transfer between the two components, which leads to its stability. The hydrogen adsorption was more favourable at HEA sites rather than that at $MoS_2$, which could explain the enhanced hydrogen gas sensing by HEA- $MoS_2$ composite.

## 2. Experimental Details

**Materials**

The Molybdenum disulphide (99%) was obtained from Thermo Fisher Scientific (USA) and used for cryomilling without any processing. Similarly, the metals purchased from Thermo Fisher Scientific (USA), Copper, Silver, Gold, Palladium, and Platinum were used to synthesize HEA NPs using arc melting, followed by cryomilling. The DMF(N,N-Dimethylformamide) 99% was also purchased from Thermo Fisher.

**Nanoparticles preparation**

Equiatomic ($Au_{0.20}Ag_{0.20}Cu_{0.20}Pd_{0.20}Pt_{0.20}$) HEA nanoparticles have been prepared using arc melting and casting followed by cryomilling method. The all five metals in equal molar ratio have been arc melted and cast as ingot in an inert atmosphere and homogenized at 1000 °C for 24 h. Subsequently, The cast ingot has been parted into smaller pieces and cryomilled until the formation of finely dispersed nanoparticles and the detailed process can be found elsewhere[49].

**Exfoliation of $MoS_2$**

The $MoS_2$ powder has also been cryomilled for 7 hours, and the sample has been collected 2, 4, and 7 hours for characterization. After 7 hours of cryomilling, the $MoS_2$ powder has been dispersed in 50:50 ratio water/DMF (Dimethylformamide) using ultrasonication. In addition, 1 wt% AgAuCuPdPt HEA nanoparticles also added and sonicated for 30 hours using ultrasonicator operated at 40 Hz, 150 watt for exfoliation of multilayers $MoS_2$.

**Characterizations**

The $MoS_2$ powders has been collected after 2, 4, and 7 hours of cryomilling are characterized using X-ray diffraction (Panalytical, Cu target, λ = 0.154026 nm). The exfoliated $MoS_2$ in 50:50% DMF/water has been washed with methanol using ultracentrifugation and dispersed in methanol. The one drop of washed and dispersed $MoS_2$-NPs was placed over carbon supported TEM grid (600 mesh Cu) and dried overnight before Transmission Electron Microscopic (TEM) analysis (FEI, Titan $G^2$ 30-600, operated at 300 kV). Similarly, one drop placed over the glass slide and dried overnight for Raman Spectroscopy analysis using 532 nm laser. The samples were drop casted on to gold coated glass slides for X-ray Photoelectron spectroscopy (XPS) and Ultraviolet photoelectron spectroscopy (UPS) measurements (Omicron ESCA 2SR XPS system equipped with Mg $K_α$ 1253.6 eV and He I 21.2 eV)

The gas sensing studies have been done in an in house-built sensor characterization system. The details of the system can be found elsewhere[50]. For measuring the sensor properties, the samples suspended in ethanol are drop-casted on the Si substrate having thermally grown oxide layer and Au interdigitated electrodes deposited by DC sputtering. The sensor current/ resistance is monitored at different temperatures in presence of various gas concentrations. The sensor response is calculated as shown in Eq. (1)

$$Sensor\ response\ \% = \frac{\Delta R}{R} \times 100 \qquad \ldots(1)$$

**Computational Methodology**

Density functional theory (DFT) was utilized for all theoretical calculations as implemented in the Vienna ab initio simulation (VASP) software (5.4.1 version)[51]. Electron-ion interactions were described using ultrasoft pseudopotentials[52]. The Perdew-Burke-Ernzerhof (PBE) method of generalized gradient approximation (GGA)[53] was used to approximate the electronic exchange and correlations. A vacuum of 15 Å was kept in the z-direction to avoid spurious interactions between the periodic images.

### 3. Results and Discussion

The $MoS_2$ nano-flakes were prepared using inhouse designed low temperature (<123 K) grinding. The layered $MoS_2$ structure was massively been fractured/exfoliated at extremely low temperature and forms 2D- $MoS_2$. Subsequently, the nanosheet of $MoS_2$ are decorated using HEA (AgAuCuPdPt) NPs allow on its surface. It is confirmed using Raman spectra of $MoS_2$ nanosheet with and without NPs as shown in Figure 1b. In case of HEA NPs decoration, Raman band has been found to be red shifted by ~2.5 cm$^{-1}$, which could possibly be due to functionalization of AgAuCuPdPt NPs over $MoS_2$. The intensity $E_{2g}$ (381 cm$^{-1}$; in-plane vibration) and $A_{1g}$ (406.7 cm$^{-1}$; out of plane vibration) of the bulk $MoS_2$ with $MoS_2$ nanoflakes after 30 hours sonication shows quite higher intensity. The process of cryomilling in successive hours (2, 4, 7 hours) has progressively led to reduction of the $MoS_2$ Raman intensity as shown in Figure S1(a), [Supplementary information] which is considered the evidence of nanostructure formation by cryomilling. In addition, the process of the 7 hours cryomilled followed by 30 hours sonication has led to drastic decrease of the intensity of $E_{2g}$ and $A_{1g}$ band as shown in Figure S1(a), which reveals the nanostructured formation with few layers of $MoS_2$. However, the addition of high entropy alloy nanoparticles (AgAuCuPdPt) in 7 hours cryomilled $MoS_2$ cause of enhancement of the intensity, indicating that the HEA alloy NPs impart surface enhance Raman scattering effects possibly due to electromagnetic field redistribution[14-15].

The X-ray diffraction (XRD) pattern of MoS$_2$ nanoflakes with NPs reveals presence of constituent phases as shown in Figure 1c. Successive hour of milling as shown in Figure S1(c) [Supplementary information] reveals that the intensity peaks in the XRD pattern continuously decreasing as time for cryomilling increases due to massive fracture multilayer MoS$_2$[16], which could result in reduced intensity in diffraction pattern as shown in inset of Figure S1(c) (002) peak. Besides, it was observed that the FWHM of the (002) increases with increasing cryomilling time in Figure S1(c).

The photo-absorption (UV-visible) of nanostructured MoS$_2$ is shown in Figure 1d, and it is found to be weak as compared to that of bulk counterparts[17-18]. It is further weakening in intensity as milling time increases as shown in Figure S1(b). The bulk MoS$_2$ after 30 hours ultrasonication reveals the highest absorbing band at 395, 450, 612 and 674 nm, which can be attributed to near band edge absorption of direct bandgap (rising edge at 700 nm), the four excitonic absorptions (612 (A) and 674 nm (B) near top of valance band and 460 (C) and 395 (D) from deep valance band respectively)[6, 19]. However, the HEA NPs in the methanol shows surfaces plasmon resonance band at nearly $\lambda_{max}$ = 284 nm. Consequently, the combination of MoS$_2$ decorated with NPs reveals the combined absorption pattern as shown in Figure1d. The cryomilled MoS$_2$ with the addition of 30 hours ultrasonication was found continuously decreasing the intensity and increasing the full width at half maxima for 2, 4 and 7 hours of milling, signifying that cryomilling leads to nanostructured MoS$_2$ with exfoliated in shorter time of ultrasonication in the solution.

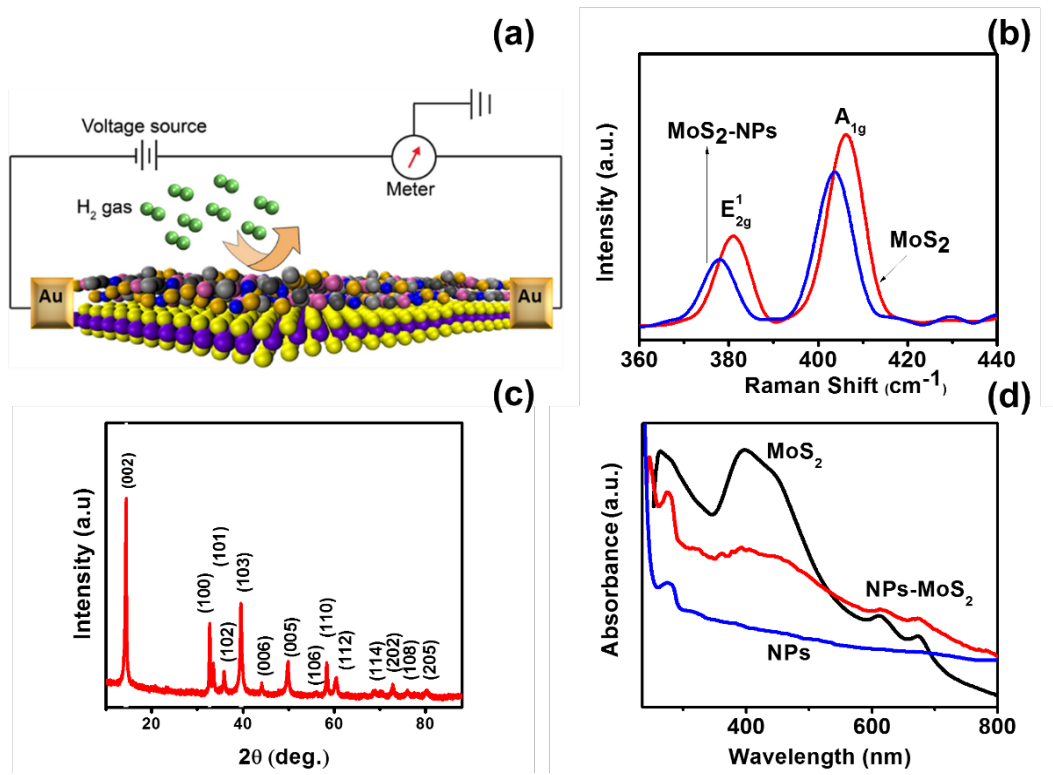

**Figure 1 (a)** Schematics **(b)** Raman spectrum of bulk MoS$_2$ without and cryomilled with AgAuCuPdPt **(c)** X-ray diffraction pattern **(d)** absorbance spectra of MoS$_2$ and AgAuCuPdPt.

The size and crystallinity of MoS$_2$ flakes functionalized or decorated with AgAuCuPdPt nanoparticles has been estimated using Transmission electron microscopy. The Figure 2(a) shows the bright field TEM image of few layer MoS$_2$ sheet, and Figure 2(b) is selected area electron diffraction pattern of monolayer or few layers MoS$_2$. The NPs were decorated with MoS$_2$ as sheen in Figure 2(c) and confirmed by electron ring diffraction pattern shown in Figure 2(d). The high resolution TEM image of single NPs is shown in Figure 2(e) and Fast Fourier fileted image of HRTEM single nanoparticles shown Figure 2(f), where the plane (111) has been marked. In addition, chemical homogeneity and composition of nanoparticles have been confirmed with TEM-EDAX mapping and energy spectra are shown in Figure 2(g-h), and the inset shows the HAADF image of nanoparticle.

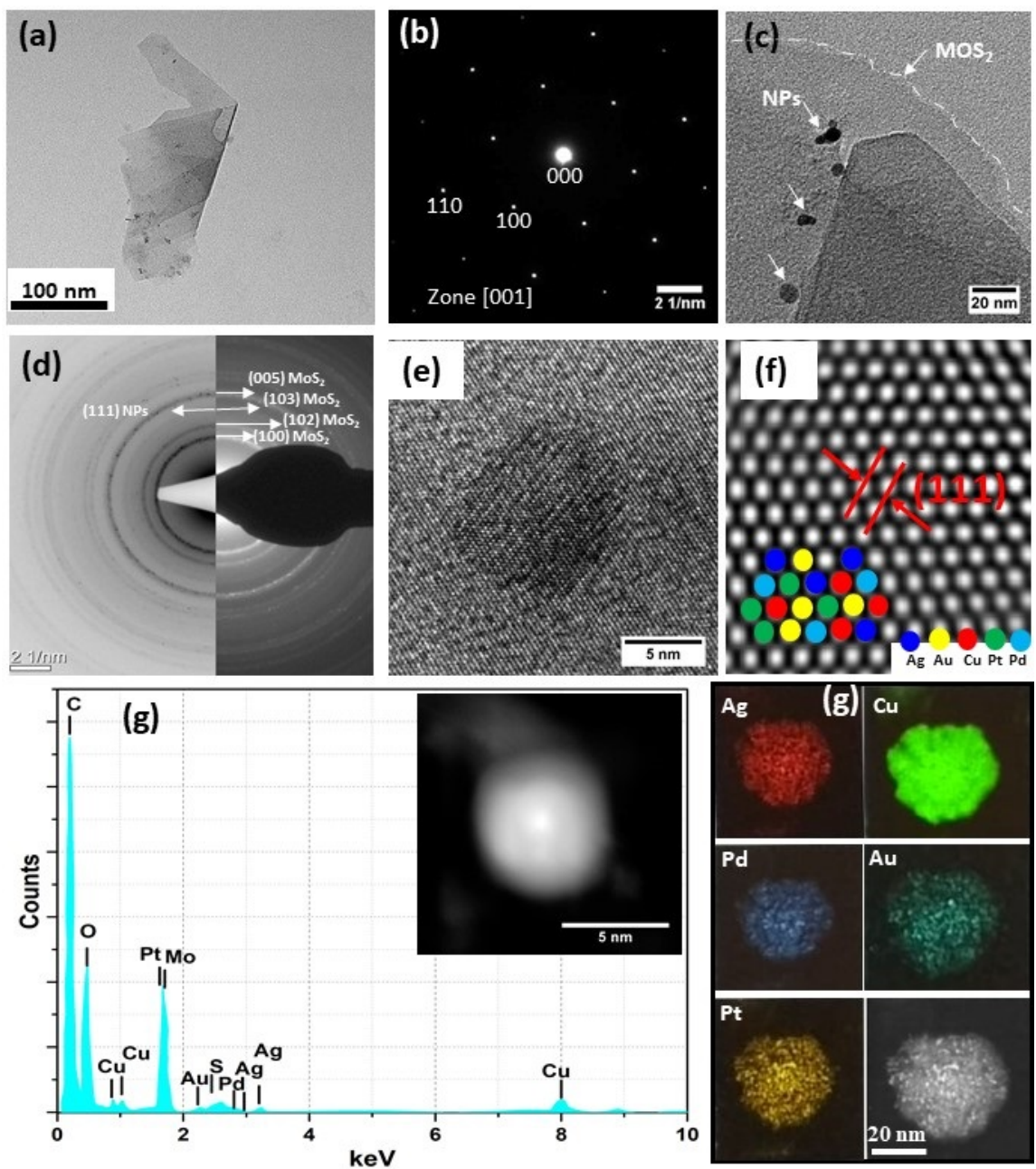

**Figure 2:** **(a)** TEM Bright field image of MoS$_2$ sheet **(b)** SAED pattern of MoS$_2$ sheet **(c)** TEM Bright field image of MoS$_2$ sheet decorated with NPs **(d)** corresponding electron ring diffraction pattern **(e)** HRTEM image of AgAuCuPdPt on MoS$_2$ sheet **(f)** FFT filtered HRTEM image of NPs **(g)** EDAX pattern inset shows single nanoparticles HAADF image **(h)** elemental mapping of NPs.

In order to study the effect of addition of HEA NPs on chemical stoichiometry and electronic structure, the X-ray and UV photoelectron spectroscopy of both the samples were conducted and the results obtained are analysed and presented in Figure 3. It is seen from Figure 3(a), that the Mo *3d* core level XPS spectra is quite complex unlike CVD grown single MoS$_2$ sheets[20]. Several peaks were observed in this region which have been attributed to core levels of S *2s* and Mo *2p* for Mo$^{4+}$ and Mo$^{6+}$ oxidation states. The peaks are resolved considering the spin orbit coupling for total angular momentum of Mo 2p 3/2 and 1/2 for each oxidation state. Ideally, for clean, stoichiometric MoS$_2$, only Mo$^{4+}$ state is expected. However, it is observed that a significantly large Mo$^{6+}$ contribution, which could be resolved clearly in either case i.e. at 229 and 232 eV for 3/2 of Mo$^{+4}$ and Mo$^{+6}$ respectively. Moreover, the relative contribution of Mo$^{6+}$ was found to be higher for HEA decorated MoS$_2$ sample. This implies that there exists surface or edge oxidation layer on the MoS$_2$ flakes which results in formation of "MoO$_3$ like" character as found in literature[21-23]. Similarly, S *2p* core level spectra shown in Figure 3(b) reveals more than one oxidation states for S (S$^{2-}$, S$_2^{2-}$) as well[24-25]. Along with this, metal sulphide peaks could also be seen at lower binding energy i.e.,159.80 eV. In addition to anionic sulphur, several peaks have been found on higher binding energy which reflect more positively charged nature of sulphur ions and hence have been attributed to oxidized sulphur i.e., SO$_x$. Hence, this supports the existence of surface oxidation with observed higher Mo$^{6+}$ ratio in the same.

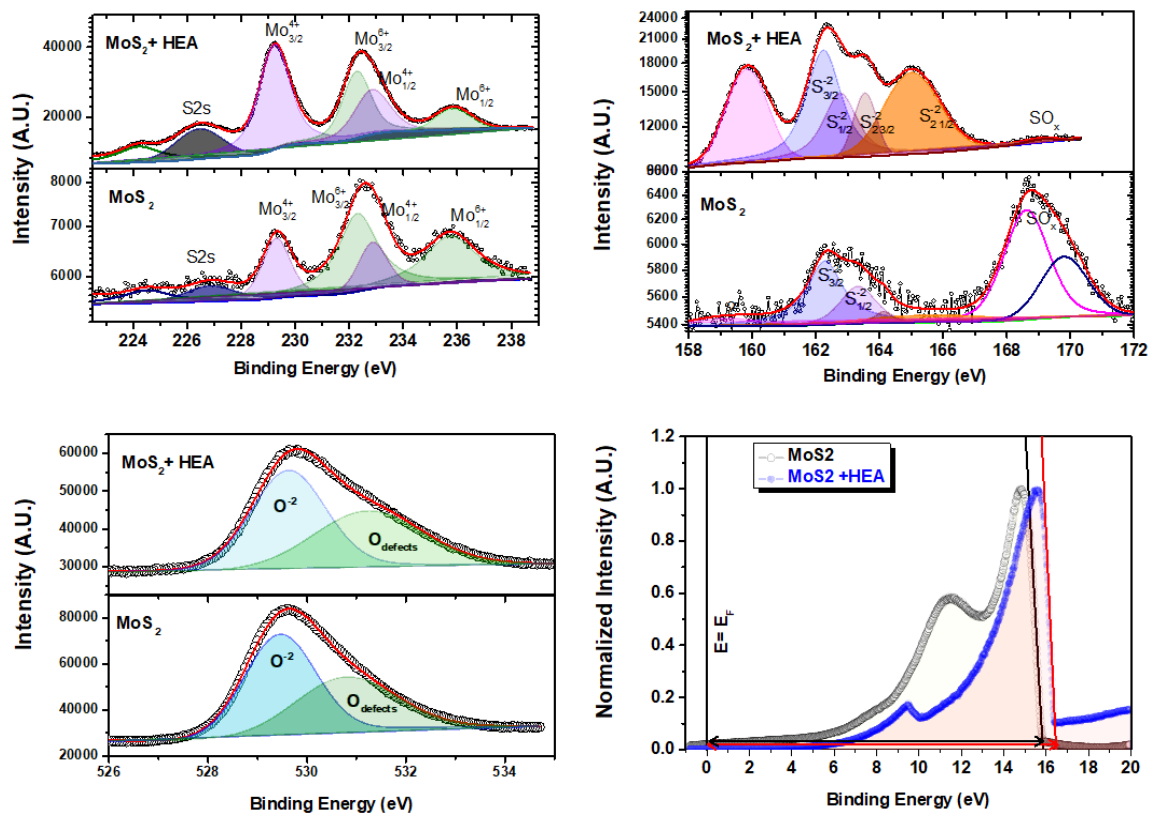

Figure 3: The X-ray photoelectron spectra of (a) Mo 3d- S 2s, (b) S 2p, (c) O 1s core levels and (d) UV photoelectron spectra of $MoS_2$ and $MoS_2$-HEA Quantum Dots.

It is known from the literature that when $MoS_2$ is made by physical methods such as sputtering, laser ablation etc. and exposed to air, it results in surface oxidation[15] due to poor crystallinity as well as rather large surface area. Thus, it is important to examine the O *1s* core level in these samples and it has been shown in Figure 3(c). As it is seen, it shows a significant oxygen peak along with a high binding energy contribution which arises because of surface defects, chemisorption etc[26]. Further, the Ultra Violet (UV) photoelectron Spectroscopy (UPS) was performed to examine the changes in the work function of the sample upon decoration with HEA NPs. As seen in Figure 3(d), as a result of HEA decoration on $MoS_2$ flakes, the secondary electron cut off is shifted towards higher binding energy which reflects a lower work function of the HEA decorated sample. The work functions calculated have also been verified using Kelvin Probe system (from KP Technology, UK) at room temperatures and the values measured are about 4.4 eV and 4.0 eV for bare and HEA decorated $MoS_2$. These values are in good agreement with those measured using UPS (photon energy – (Secondary electron cut off – Fermi energy). Thus, it is evident that the HEA nanoparticles affect the electronic transport of the material which may result in forming a schottky barrier.

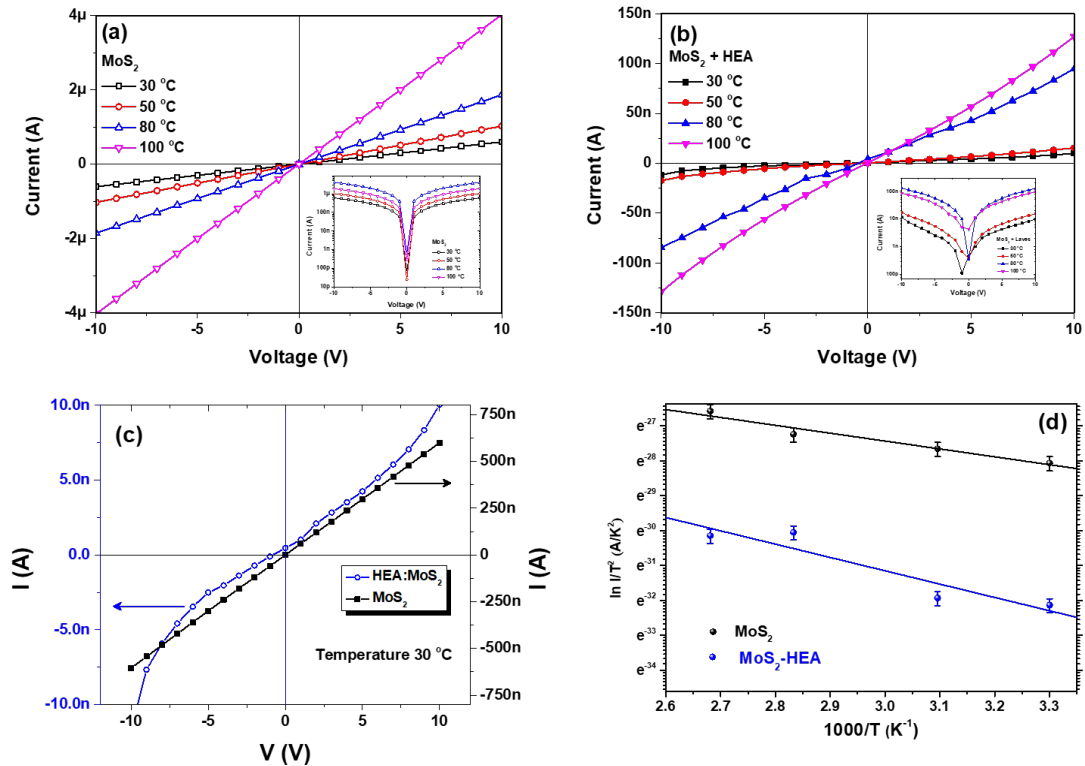

Figure 4: The current-voltage (I-V) characteristics of devices fabricated using (a) only $MoS_2$ and (b) HEA NPs decorated $MoS_2$ as a function of temperature. (c) the comparison of linearity and current of both devices at 30 OC. (d) the Arrhenius plot of ln ($I/T^2$) vs 1000/T for both the samples.

Figure 4(a) and 4(b) shows the current–voltage characteristics of pristine $MoS_2$ and HEA:$MoS_2$ respectively as a function of temperature. As seen from Fig 4(c) the devices made using Au electrodes show a linear I-V characteristic for pristine $MoS_2$, as expected[27-28]. On the other hand, the HEA:$MoS_2$ shows a nonlinear I-V characteristic indicating a non-ohmic i.e. schottky nature accompanied with lower conductance of the device. This could be due to change in the chemical potential of $MoS_2$ upon decoration with HEA NPs. However, both the devices show semiconducting nature, i.e. the instantaneous resistance decreases upon heating. Thus, in order to extract the barrier height between at the metal-semiconductor junction in either case, the I-V is analysed for thermionic emission mechanism as it is dominant in absence of any gate voltage (two probe device). Here, the thermionic emission current is given by the following Richardson –Dushman equation (2)[29-30]

$$I = SA^*T^2 \, e^{-\frac{q\phi}{kT}} [1 - e^{\frac{qV}{kT}}] \qquad (2)$$

Where, I is the current obtained at applied bias V, T is absolute temperature, q is the charge of electron, $k$ is Boltzmann constant, $A^*$ is Richardson constant and $S$ is the area of the device. The same equation (2) may be written in the form of equation (3) as

$$\ln(I/T^2) = \ln C - \frac{q\phi}{kT}, \qquad (3)$$

Where, $C = SA^* \, [1 - e^{\frac{qV}{kT}}] \qquad (4)$

Thus, from Eq (3) when $\ln(I/T^2)$ is plotted as a function of 1/T, the slope gives the value of barrier height $\phi = |\Phi_{metal} - \Phi_{semiconductor}|$. Ideally, the slope value obtained is further plotted against the value of potential applied to get the y-intercept i.e. in the limit of zero bias[30-31]. However, we have observed that the slope is practically independent of bias (See figure S2 in supplementary information) i.e. the barrier is sufficiently high and the change in barrier with application of electric field is minimal. This is also evident from the three orders of magnitude lower current in comparison to that of the ohmic contact formed without HEA nanoparticles.

Thus the barrier height, $\phi$ value obtained from Figure 4(d) are 0.195 and 0.35 eVs for bare $MoS_2$ and HEA decorated $MoS_2$, respectively. Considering the work function of gold, 5.1 eV, the work function values deduced for bare and HEA decorated $MoS_2$ are 4.9 and 4.75 eV respectively. These values are in good agreement with those obtained using UPS measurements and KP measurements. This implies that upon decoration with HEA nanoparticles, the work function of $MoS_2$ has decreased as already seen from Figure 3(d).

The gas sensing response of the devices was studied for various concentrations of hydrogen gas. As can be seen from Figure 5(a), the bare $MoS_2$ without HEA NPs shows fairly low response to 1000 ppm hydrogen gas at room temperature (i.e. 30°C). Nonetheless, the HEA:$MoS_2$ shows a nearly 10 fold increase in the sensor response when heated to moderate temperatures like 50-80°C. The bare $MoS_2$ shows consistently low response at all temperatures while, the HEA:$MoS_2$ sensor response shows a maxima at 80°C and drops when heated beyond. Further, when the response is studied towards various concentrations of hydrogen at various temperatures, it is found to obey power law of semiconductor gas sensors(equation 5)[32-33], as shown in Figure 5(b).

$$S = AC^\beta \tag{5}$$

Here, this equation can be deduced from Fruedlich adsorption isotherm wherein A is a prefactor governing the adsorption kinetics and n is the exponent which depends on surface coverage[34]. Moreover, the HEA NPs when decorated over $MoS_2$ not only enhance response but also showed improved response time and recovery times. For instance, the comparison of the response and recovery time for pristine $MoS_2$ and HEA:$MoS_2$ is shown in Figure 5(c). Here, the response and recovery times are calculated as time required to reach the 90% of the saturation value and 90% under the saturation value respectively. It may be observed from Figure 4(c) that the sensors of pristine $MoS_2$ and HEA:$MoS_2$ take 10 min and 15 min as response time while 7 min and 16 min as recovery times respectively.

The Faster response may be due to catalytic activity of the HEA NPs which helps in improving the number as well as rate of the surface adsorption of the target gas molecules. For practical applications it is ideal that the sensor responds and recovers faster as shown in Figure 5(d) for various concentrations of hydrogen. These obtained time constants with high response are much smaller than those reported to date to the best of our knowledge[9, 30].

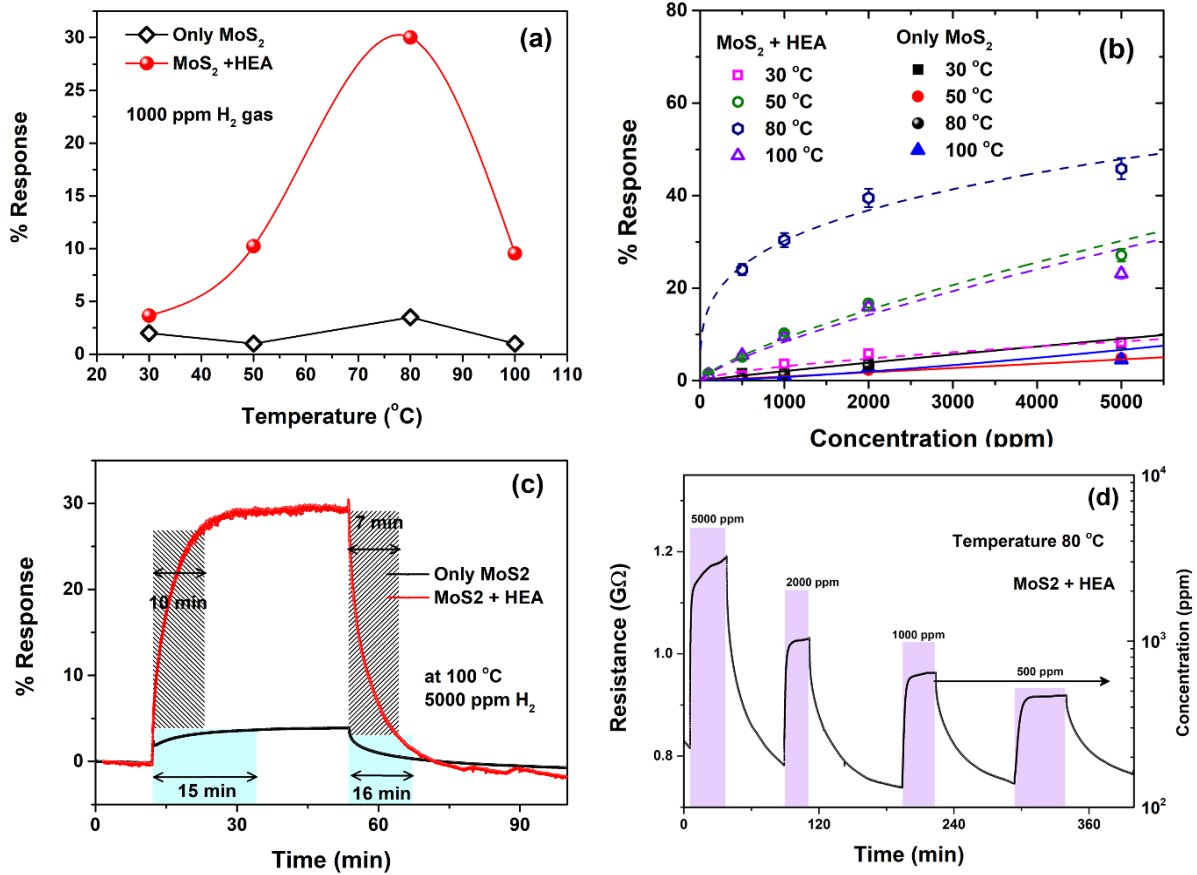

Fig.5 The comparison of sensor response only MoS$_2$ and HEA NPs decorated MoS$_2$ with (a) temperature and (b) hydrogen gas concentration. The lines show power law fit. (c) The comparison of response time and recovery times for 5000 ppm H$_2$ gas at 100°C. (d) The typical response of HEA NPs decorated MoS$_2$ at 80°C for various concentrations.

Hydrogen is a reducing gas, i.e. it donates an electron when it surface adsorbs. This electron when received in an n-type semiconductor, leads to increase in its carrier concentration and hence resistance drops. While in p-type semiconductors, it recombines with some of the holes and thus decreases the carrier concentration and hence resistance increases. The rise in resistance observed in this study denotes that the observed behaviour is that of the p-type semiconductor and this p-type behaviour was attributed to the surface oxygen adsorption on MoS$_2$ flakes[35]. Further, the type of carriers is also confirmed by hall measurements. The lower work function implies that the fermi level (zero binding energy line) moving closer to conduction band (away from valance band), which will make the barrier height more for n-type material. On the other hand, the fermi level moving closer to valance band will make the barrier height higher for p-type semiconductor.

High entropy alloys are solid solutions of five metal atoms which are otherwise highly catalytically active materials. Whenever such catalytically active metal atoms are attached to semiconductor gas sensor material, it is known to enhance the sensor response by two mechanisms, i.e. controlling the fermi level and spill over mechanism. The fermi level control results due to changes in electronic transport as a result of formation of a depletion region, while the spill over refers to the catalytic effect of these particles[36-38]. The spill over model is described as enhanced adsorption of gaseous species on metals and self-transportation of these species on to the supporting sensor material[39]. Here, the support material does benefit from additional gas species adsorption which it does not achieve otherwise. The noble metal elements, particularly Pd[25] and alloys of noble metals enhance the surface adsorption for hydrogen due to such dissociative adsorption of hydrogen atoms[38, 40-41] followed by transfer of adsorbed gas molecules/ atoms to the support materials i.e. $MoS_2$ in this case. This transfer further results in improved charge transfer and hence better catalysis or sensor response. On the other hand, as seen from Figure 4, the electronic levels of the semiconductor are affected upon decoration with metal nanoparticles. At the junction of metal and semiconductor, there exists a barrier due to charge transfer. Metals being strong donor of electrons, these are pumped into the semiconductor. Particularly in case of p-type semiconductors, these electrons recombine with some of the holes which are majority charge carriers and hence result in restoration of Fermi level towards the intrinsic limit. This increases the resistance of semiconductor as seen in figure 4. Further, the formation of nanoscale barriers at the HEA alloy and $MoS_2$ junctions is beneficial for the sensing as it produces a large change in resistance when there is a small change in carrier concentration of $MoS_2$ due to hydrogen chemisorption.

Thus, the synergistic effect of all five metal species leads to a giant ten-fold increase in sensor response. Further, the kinetics of surface adsorption decides the sensor response time taken for saturation and recovery. Here, the strong affinity of HEA NPs towards hydrogen results into faster response as well as recovery of the sensor signal at much lower temperatures than those reported in the literature so far[25]. Moreover, the surface of $MoS_2$ which is rich in oxygen species after decorating with HEA, further enhances the abundance of active sites for surface adsorption and making it more sensitive.

To gain atomistic insights into the enhanced gas sensing at the $MoS_2$-HEA system, density functional theory (DFT) calculations were performed. The optimized lattice parameters of 2D $MoS_2$ and pristine Pt were found to be: a = b = 3.19 Å, and a = b = c = 3.96 Å, in excellent agreement with previous reports[42-43]. The AgAuCuPdPt alloy (HEA) special quasirandom structure (SQS) was generated from 5x2x1 supercell of Pt (111) surface using the "mcqs" code employed in Alloy Theoretic Automated Toolkit (ATAT)[44]. The ATAT code employs a Monte Carlo simulated annealing method using an objective function that perfectly matches the maximum number of correlation functions[44]. The SQS generated in such a way mimics the thermodynamic and electronic properties of true structures to a reasonable extent, as has been shown in the earlier reports[45-46]. The unit cell of the SQS (HEA) contained 80 atoms, with the Pt, Pd, Ag, Au and Cu atoms comprising 20% each of the total composition. The 9x4x1 supercell of $MoS_2$ was matched with this unit cell of HEA, resulting in a lattice mismatch of 2.4%.

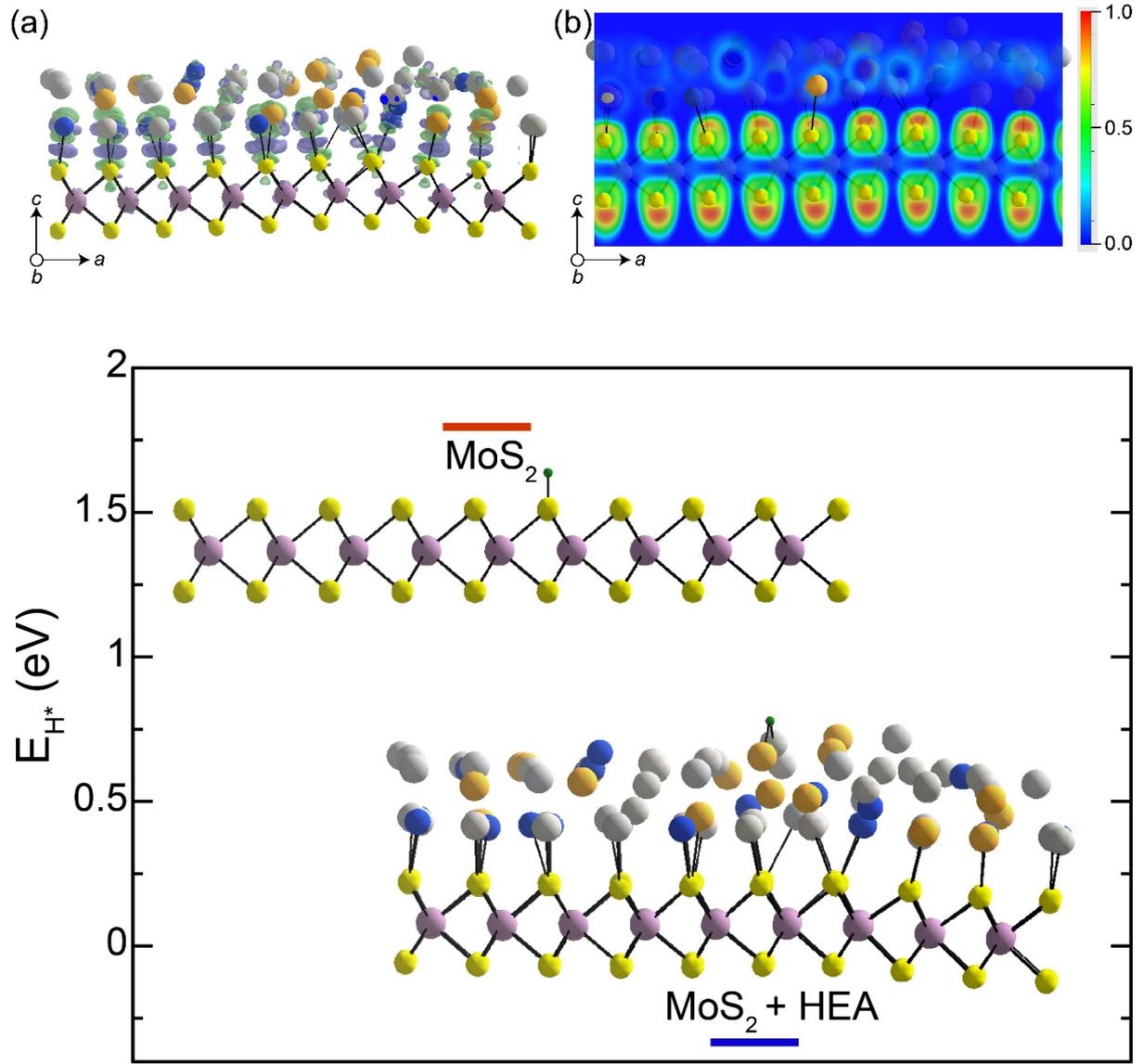

*Figure 6.* *(a)* Charge density difference plotted for $MoS_2$-HEA system. The charge accumulation and depletion regions are indicated by lime green and dark violet isosurfaces, respectively. The isosurface value has been set to $4 \times 10^{-3}$ eÅ$^{-3}$. *(b)* ELF plot for $MoS_2$-HEA system along (010) plane. The color bar for ELF values are also shown alongside. The light violet, yellow, light grey, dark grey, gold, silver, blue, and green spheres represent Mo, S, Pt, Pd, Au, Ag, Cu, and H atoms, respectively. *(c)* $E_{H^*}$ of $MoS_2$ (red solid line) and $MoS_2$-HEA (blue solid line) systems. The completely optimized H-adsorbed structures are also shown where light violet, yellow, light grey, dark grey, gold, silver, blue, and green spheres represent Mo, S, Pt, Pd, Au, Ag, Cu, and H atoms, respectively.

To check the thermodynamic stability of the $MoS_2$-HEA composite system, interface binding energy was calculated according to the expression [47]:

$$E_b = (E_{MoS_2+HEA} - E_{MoS_2} - E_{HEA}) / A \qquad \_\_(1)$$

Where $E_{MoS_2+HEA}$, $E_{MoS_2}$, and $E_{HEA}$ are the total energies of MoS₂-HEA composite, MoS₂, and HEA, and $A$ is the area of the unit cell (Table S1). The value for $E_b$ was found to be -74.02 meV/Å², indicating its high stability, in agreement with the experiments. This is further corroborated by the charge density difference ($\rho_{diff}$) plot shown in Figure 6(a), which was calculated using:

$$\rho_{diff} = \rho_{MoS_2+HEA} - \rho_{MoS_2} - \rho_{HEA} \quad \underline{\quad} (2)$$

Where $\rho_{MoS_2+HEA}$, $\rho_{MoS_2}$, and $\rho_{HEA}$ are charge densities of the MoS₂-HEA composite, MoS₂, and HEA, respectively. A significant charge transfer is observed at the interface of MoS₂ and HEA, which indicates large interaction between the two layers and hence high stability of the composite. To analyse the behaviour of bonding in the composite, electron localization function (ELF) was calculated [48]. The covalent (ELF > 0.5) and metallic bonding (ELF < 0.5) nature of MoS₂ and HEA layers, respectively, are retained after formation of the composite (Figure 6(b), and Figure S3). However, the red regions localized at S atoms (corresponding to their lone pairs) at the interface of MoS₂ and HEA are non-uniform, i.e., its intensity is low where atoms of HEA are closer to MoS₂. It again signifies an appreciable interaction strength between the two layers.

Next, in order to explain the experimental trends observed for response time, H adsorption energy calculations were performed according to the expression given below:

$$E_{H^*} = (E_{tot} - E_{pristine} - \frac{1}{2}E_{H_2}) / A \quad \underline{\quad} (3)$$

where $E_{tot}$, $E_{pristine}$, and $E_{H_2}$ are the total energies of H adsorbed system, pristine (bare) system, and isolated H₂ gas molecule, respectively (Table S1). The values for $E_{H^*}$ corresponding to MoS₂ and MoS₂-HEA composite are shown in Figure 1, where it can be seen that $E_{H^*}$ on MoS₂-HEA composite is exothermic, while that on MoS₂ is endothermic.

Therefore, H will tend to adsorb faster on MoS$_2$-HEA composite compared to that on bare MoS$_2$, hence the response time will be lower for the MoS$_2$-HEA composite, as is also observed experimentally.

## 4. Conclusion

To summarise, we have proposed a facile method to decorate/functionalise 2-D materials. Further, we have investigated its effect on gas sensing properties by comparing with bare MoS$_2$ that accounts for unusual p-type conduction. A powerful technique for chemical compositional analysis (XPS) has been carried out to account for the unusual p-type conduction in MoS$_2$. It is evident that both bare MoS$_2$ and functionalised MoS$_2$ can be operated at lower temperature unlike MOS gas sensors. The calculated values of $E_{H*}$ from DFT for MoS$_2$ and functionalised MoS$_2$ are in good agreement with the experiment and proves that the HEA-functionalised MoS$_2$ is effective in achieving higher sensitivity than bare MoS$_2$ with comparatively lower response and recovery times. The non-linear nature of contact between gold and HEA decorated MoS$_2$ reveals the potential for enhancement in MoS$_2$ gas sensing properties.

**Acknowledgment**
KU and VK would like to thank the DST for the Inspire faculty award grant and IISER TVM central instrumental facility for the XPS-UPS measurements. R.K and A.K.S acknowledge the computational facilities located at Materials Research Centre (MRC), SERC and Solid State Structural and Chemistry Unit (SSCU), at Indian Institute of Science, Bangalore and DST-Korea for funding.

# Supplementary Information

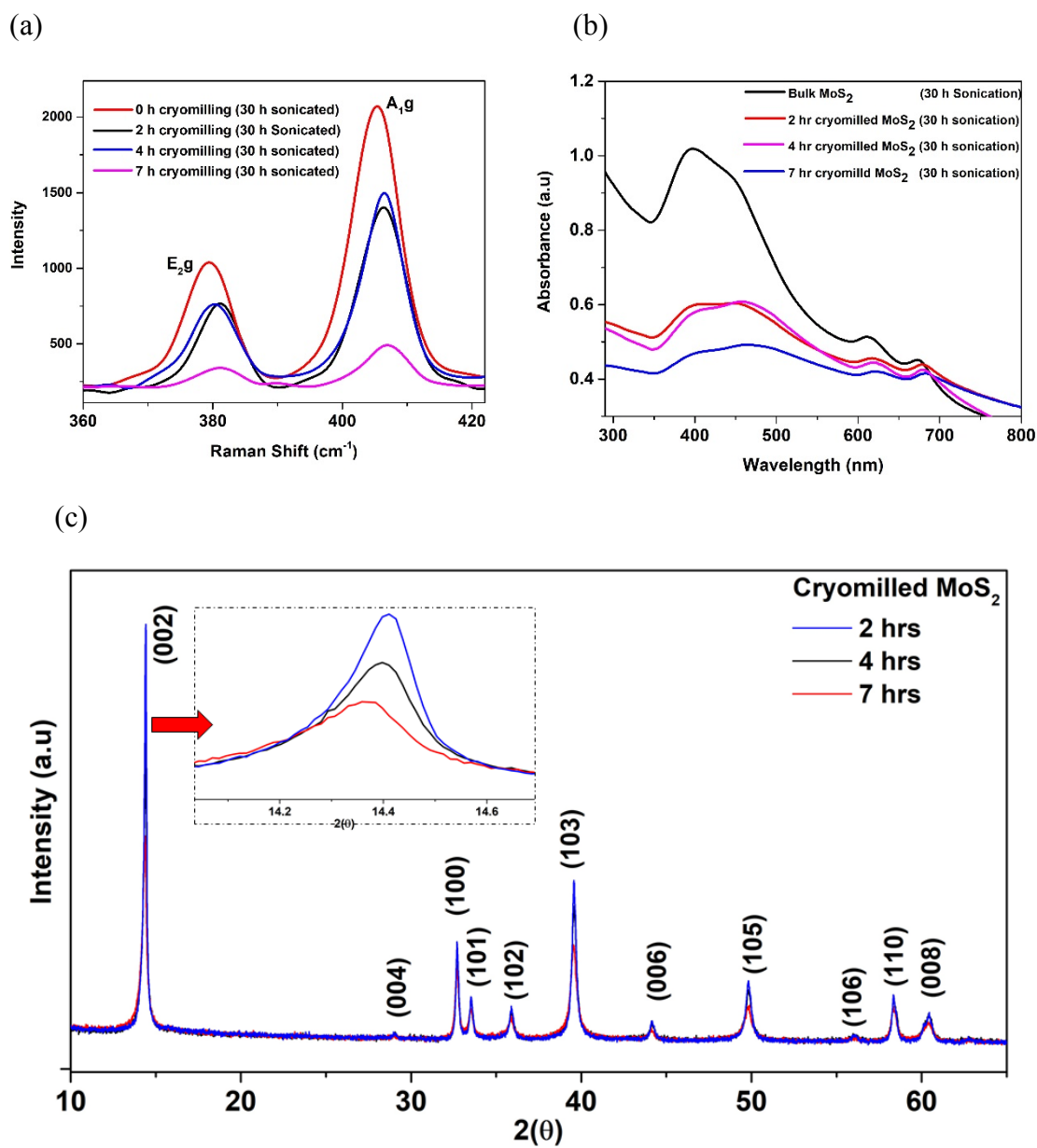

*Figure S1:* **(a)** Raman spectrum of successive hours cryomilled and sonicated for 30 hours **(b)** absorption spectra of $MoS_2$ in successive hours cryomilling **(c)** X-ray diffraction pattern of successive hours cryomilling of $MoS_2$.

**Hall Effect:**

Hall effect measurement was done at two different values of Magnetic field i.e. 2000 G and 4000 G to confirm the carrier type and it confirms the majority carriers as **p-type.**

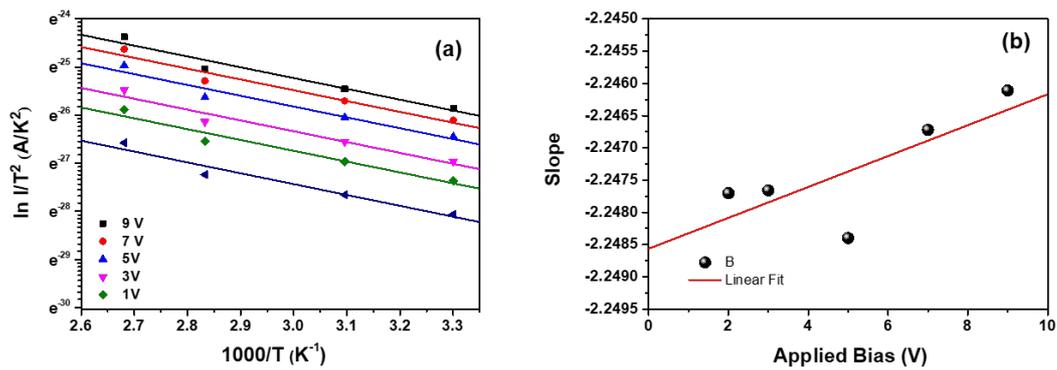

**Figure *S2* (a)** The variation of I/T² Vs 1/T at various applied biases and of slope at each bias voltage.